# SumVg: Total heritability explained by all variants in genome-wide association studies based on summary statistics with standard error estimates

**Hon-Cheong SO[1-7], Xiao XUE[1], Pak-Chung SHAM[8]**

[1]School of Biomedical Sciences, The Chinese University of Hong Kong, Shatin, Hong Kong

[2]KIZ-CUHK Joint Laboratory of Bioresources and Molecular Research of Common Diseases, Kunming Institute of Zoology and The Chinese University of Hong Kong, China

[3] Department of Psychiatry, The Chinese University of Hong Kong, Hong Kong

[4] CUHK Shenzhen Research Institute, Shenzhen, China

[5]Margaret K.L. Cheung Research Centre for Management of Parkinsonism, The Chinese University of Hong Kong, Shatin, Hong Kong

[6]Hong Kong Branch of the Chinese Academy of Sciences Center for Excellence in Animal Evolution and Genetics, The Chinese University of Hong Kong, Hong Kong SAR, China

[7]Brain and Mind Institute, The Chinese University of Hong Kong, Hong Kong SAR, China

[8] Department of Psychiatry, The University of Hong Kong, Hong Kong

**Abstract:**

Genome-wide association studies (GWAS) are commonly employed to study the genetic basis of complex traits/diseases, and a key question is how much heritability could be explained by all variants in GWAS. One widely used approach that relies on summary statistics only is LD score regression (LDSC), however the approach requires certain assumptions on the SNP effects (all SNPs contribute to heritability and each SNP contributes equal variance). More flexible modeling methods may be useful.

We previously developed an approach recovering the "*true*" *z*-statistics from a set of *observed z*-statistics with an empirical Bayes approach, using only summary statistics. However, methods for standard error (SE) estimation are not available yet, limiting the interpretation of results and applicability of the approach.

In this study we developed several resampling-based approaches to estimate the SE of SNP-based heritability, including two jackknife and three parametric bootstrap methods. Simulations showed that delete-*d*-jackknife and parametric bootstrap approaches provide good estimates of the SE. Particularly, the parametric bootstrap approaches yield the lowest root-mean-squared-error (RMSE) of the true SE. In addition, we applied our method to estimate SNP-based heritability of 12 immune-related traits (levels of cytokines and growth factors) to shed light on their genetic architecture.

We also implemented the methods to compute the sum of heritability explained and the corresponding SE in an R package SumVg, available at https://github.com/lab-hcso/Estimating-SE-of-total-heritability/. In conclusion, SumVg may provide a useful alternative tool for SNP heritability and SE estimates, which does not rely on distributional assumptions of SNP effects.

**Introduction**

Genome-wide association studies (GWAS) have proven to be successful in dissecting the genetic basis of a variety of diseases. A number of new susceptibility loci have been discovered, providing novel insight into the pathophysiology of many diseases. Nevertheless, a large proportion of the heritability still remained unexplained. It is natural to question the maximum variance that could be explained by all variants in a GWAS (or meta-analyses of GWAS), as we expect many true susceptibility variants are "hidden" due to limited power.

A number of methods have been developed to estimate the total heritability by all measured SNPs (also known as SNP-based heritability). Regarding methods that require individual-level data, in a pioneering work, Yang et al [1] derived a method to estimate the variance explained by all SNPs in a GWAS by a liner mixed model with random SNP effects. The approach assumes all SNPs have non-zero and normally distributed effects (beta), with a mean effect of zero. Each SNP is assumed to contribute to the same level of explained variance (i.e., variance explained by each SNP = total heritability/number of SNPs). Other similar approaches have also been proposed. For example, LDAK [2] assumes different heritability explained each SNP, depending on the minor allele frequencies (MAF), LD score and imputation quality of the SNP. Advanced methods have also been developed to estimate SNP-based heritability using summary statistics alone. LD score regression (LDSC) is one of the most widely used approaches for this purpose [3]. LDSC assumes a mean effect (beta) of zero and equal variance explained by each SNP (i.e., an infinitesimal model). SumHer [4] is an alternative approach based on the LDAK assumptions. For a more detailed review, please refer to [5].

Prior to the development of LDSC, we have developed an alternative framework ([6]; referred to as "SumVg" in this paper) to achieve the same goal of estimating SNP-based heritability using summary statistics alone. Essentially, we aimed to recover the "*true*" *z*-statistic from a set of *observed* *z*-statistics based the following formula established by Brown [7] and Efron [8]. The corrected *z*-statistics are then converted to variance explained. There are several advantages of this method. Most importantly, the SumVg approach does *not* rely on any distributional assumptions of the effect sizes of susceptibility variants. In addition, it does *not* assume an equal heritability explained by each SNP, or that all SNPs contribute to the heritability (infinitesimal model). There are also no assumptions on the relationship between allele frequencies and variance explained. The method is also computationally fast. Also, since the LDSC method directly leverages LD patterns, a well-matched LD reference panel is usually required [9]. There is less reliance on LD information for SumVg as LD is mainly used for pruning.

Our method has been applied in a number of studies [for example see [10-18]]. However, there are no available methods to quantify the variance or precision of the heritability estimates from SumVg. If raw data is available, a standard non-parametric bootstrap by sampling individuals with replacement could be employed. However, in most cases only summary statistics are available and there are currently no methods for evaluating the standard error of the point estimate of heritability.

We summarize the contributions of this study below. In this work, we proposed five re-sampling approaches to estimate the SE of the total heritability by all SNPs in GWAS, based on summary statistics alone. Extensive simulations were performed to compare and validate the performance of different methods. Secondly, we also presented an easy-to-use R program to implement the SumVg approach with different flexible modeling options. Thirdly, we reported heritability estimates for 12 immune-related traits (levels of cytokines and growth factors)[19] based on this approach, for which LDSC was unable to provide reasonable estimates. Such cytokines/growth factors are regulators of immune responses and inflammation, and are important intermediate phenotypes for autoimmune, inflammatory and infectious diseases[20]. As such, it is of scientific and clinical importance to unravel the genetic architecture of these traits, and estimating their heritability may be considered a useful contribution in its own right.

**Methods**

*Estimation of the total heritability explained*

Readers may refer to our previous paper [6] for details of estimation of the sum of heritability explained. In brief, we estimated the "true" z-statistics by the following correction formula:

$$E\{\delta|z\} = z + \frac{f'(z)}{f(z)} \quad \text{---------------------}(1)$$

where $z$ denotes the observed $z$-statistic and $\delta$ denotes the "true" $z$-statistic (i.e. the z-statistic one would obtain if there were no random noise; it reflects the actual effect size). Note that the method does not require prior assumptions on the underlying distributions of the true effect sizes $\delta$.

We also proposed an alternative approach by evaluating the expected effect size conditioned on $H_1$ (i.e. $\delta \neq 0$)

$$E(\delta|z, H_1) = E_1(\delta|z) = \frac{E(\delta|z)}{\Pr(H_1|z)} = \frac{E(\delta|z)}{1 - \text{fdr}(z)} \quad \text{---------------------}(2)$$

where fdr is the local false discovery rate described in Efron [21]. The resulting estimate of the true Vg is then $(1 - \text{fdr}(z))$ multiplied by the estimator in (2) converted to Vg scale.

The conditional estimator however is prone to large random variations as it involves local fdr estimation of each SNP. In many subsequent applications of our heritability estimation method [10-12], the unconditional estimator [equation (1)] was primarily employed. We shall hence focus on the unconditional estimator in this paper, although the resampling approaches described below can readily be applied to other estimators in our previous work [6] as well.

*Standard and delete-d-jackknife*

In standard (delete-one) jackknife procedure [22], we estimate the standard error (SE) by leaving out one observation at a time. The SE is defined by

$$\hat{se}_{jack} = \sqrt{\frac{n-1}{n} \sum \left(\hat{\theta}_{(i)} - \hat{\theta}_{(.)}\right)^2}$$

where $n$ is the sample size, $\hat{\theta}_{(i)}$ is the parameter estimate from the sample with the $i$ th observation removed and

$$\hat{\theta}_{(.)} = \sum_{i=1}^{n} \hat{\theta}_{(i)} / n$$

In our case the parameter is the sum of heritability from all variants.

An extension is the delete-d-jackknife [23] where we leave out $d$ observations at a time. There are in total $N = \binom{n}{d}$ possibilities of removing $d$ out of $n$ observations. In practice, $N$ is usually very large. One may simply randomly repeat the procedure $m$ times only ($m \leq N$) instead of exhausting all possibilities of removing $d$ out of $n$ observations. The standard error is given by

$$se_{del-d-jack} = \sqrt{\frac{n-d}{dm}\sum_{v=1}^{m}\left(\hat{\theta}_{S_v} - \frac{1}{m}\sum_{v=1}^{m}\hat{\theta}_{S_v}\right)^2}$$

where $\hat{\theta}_{S_v}$ denotes the parameter estimate in the $v^{th}$ jackknife replicate where $d$ observations are left out. The delete-$d$-jackknife (when $d>1$) works better than the standard jackknife for non-smooth parameters like the median [23].

There are no clear rules on the choice of $d$ in delete-$d$-bootstrap. Chatterjee [24] suggested $n/5$ as a reasonable choice for $d$ based on consideration of efficiency and likely model conditions. We followed the suggestion by Chattejee [24] and set $d$ as $n/5$ (=20000) in all simulations.

*Bootstrap approaches*
Parametric bootstrap

In parametric bootstrap, in each replication we simulated z-statistics based on $\hat{\delta}$, the 'corrected' z-statistics from original sample. We have

$$z_{i,b} \sim N(\hat{\delta}_i, 1)$$

where $z_{i,b}$ denotes the $i$ th z-statistic in the $b$ th bootstrap replicate. For small effects, the $\hat{\delta}$ will be shrunken towards zero.

We further proposed a modified approach by also considering the local fdr (i.e., probably of null given z) of each z-statistic. In each replicate, we simulated z-statistics according to the following scheme:

$$z_{i,b} \sim N(\hat{\delta}_i, 1) \text{ with a probability of } 1\text{-}\hat{fdr}(z_i)$$

$$z_{i,b} \sim N(0,1) \text{ with a probability of } \hat{fdr}(z_i)$$

Alternatively, one may employ the original z-statistics instead of the corrected z-statistics as the mean in each simulation, i.e.

$$z_{i,b} \sim N(\hat{z}_i, 1) \text{ with a probability of } 1\text{-}\hat{fdr}(z_i)$$

$$z_{i,b} \sim N(0,1) \text{ with a probability of } \hat{fdr}(z_i)$$

The standard error is then computed from the simulated z-statistics.

*Tests of resampling-based SE estimates*
We compare the SE estimated from the above methods with the 'true' SE obtained from one hundred simulations with known data generating distributions. The details of the simulations follows [6]. Briefly, a gamma distribution was used to simulate three levels of variance explained (Vg =0.101, 0.191, 0.295), which were converted to true effect sizes ($\delta$). Z-statistics for 100,000 independent SNPs (0.5% were non-null) with different sample sizes (N=5000, 10000, 20000, 50000, 100000, 200000) were then simulated as input for SumVg following the distribution $N(\delta,1)$.

Two hundred replicates were run for each bootstrap or jackknife procedure. We focus on quantitative traits in our simulations, but the results should most likely apply to binary traits as well, as the only difference in these two scenarios is the formula to convert $z$ to variance explained (Vg).

**Application to immune traits**

A selected set of immune-related traits (levels of cytokines/growth factors) were included for study, based on the GWAS by Ahola-Olli et al.[19]. We selected 12 continuous immune traits with (1) sample size $N$>5000 and (2) very low (<=3%) or negative SNP-based heritability estimated by LDSC. The LDSC heritability were based on pre-calculated values from GWASAtlas (https://atlas.ctglab.nl/). SNPs in strong LD were removed using the PLINK command "--indep-pairwise 100 25 r2" with a series of $r^2$ thresholds (0.1, 0.05, 0.025, 0.01, 0.005, 0.002, 0.001). The 1000G Phase3 EUR sample was used as the reference panel to calculate LD among variants. Independent SNPs with MAF>0.01 were then applied to SumVg.

**Results**

**Simulation results for SE estimation**

Standard errors (SE) of heritability, as estimated by jackknife and bootstrap approaches, are listed in Table 1 and plotted in Figure 1. Bias, variance, and root mean square error (RMSE) of SE were calculated over 100 simulations (Table 2; Figures 1-2).

The delete-[$n$/5]-jackknife worked reasonably well when the total heritability explained is low (when heritability=0.101), but it tended to overestimate the SE when the total heritability is higher, especially at larger sample sizes. The bias was also positive across all simulation scenarios. The standard (delete-1) jackknife approach performed the worst among all methods, producing inflated estimates of SE. The variance and RMSE of this estimator were high compared to other approaches. The SE was in general over-estimated at all heritability levels across all sample sizes. This may be explained by the fact that sum of Vg is not a very smooth parameter, which impairs the performance of delete-1-jackknife estimators.

The other methods, including parametric bootstrap (paraboot) and the modified versions with consideration of local fdr, performed reasonably well and closely resembled the true SE. With the exception of one simulation setting, the parametric bootstrap methods achieved the lowest (absolute) bias. For variance and RMSE, parametric bootstrap also performed the best. In terms of RMSE, the parametric bootstrap approaches that also model the local fdr outperformed other methods. The RMSE of different estimators were also observed to reduce with increasing sample sizes.

**Results on immune traits**

Plink was applied to trim GWAS data for 12 immunological traits at various $r^2$ criteria to obtain roughly independent SNPs. We only included common variants with MAF>0.01 for further analysis. Then, using SumVg, "true" z-statistics of trimmed SNPs were retrieved to capture the missing heritability. The jackknife and bootstrap methods were used to compute the corresponding SE (Table 4).

The total SNP-based heritabilities predicted by SumVg for the selected traits, in contrast to the comparatively low or negative heritability estimates from LDSC, were around 10-20% based on a collection of LD-pruned SNPs. We obtained a stable (and likely conservative) estimate of heritability at $r^2 \sim 0.01$ or 0.005. Lower $r^2$ values (i.e., $r^2$<0.0025 and $r^2$<0.001) had limited impact on final estimates of heritability. The del-1 jackknife consistently produced the highest standard error, while the bootstrap and del-d jackknife approaches produced SE that were more comparable to one another. Out of the 12 cytokines/growth factors study, highest heritability was observed for IL-4 and IL-17 levels.

**R package implementation**

We also implemented the methods to compute the sum of heritability explained and the corresponding SE in an R package SumVg, available at https://github.com/lab-hcso/Estimating-SE-of-total-heritability/.

**Discussions**

Here we presented an approach for estimating SE of SNP-based heritability estimates using SumVg, and our applications demonstrate the usefulness of the approach.

Our main purpose is to provide an alternative approach for SNP-based heritability and SE estimation, since different approaches have different statistical modeling assumptions, or assumptions on the genetic architecture. In practice, it is almost impossible to know the true genetic architecture of a disease/trait, and as such it is very difficult to verify the correctness of heritability estimates due to lack of a 'gold standard'. It will be more reassuring if one observes similar heritability estimates from diverse methods. SumVg may provide a useful alternative reference for heritability estimates, in conjunction with existing approaches such as LDSC. SumVg may also be useful when standard approaches are unable to give reasonable results (e.g. close to zero heritability for traits that are likely heritable from previous studies, or negative estimates).

We recommended pruning of SNPs (such that SNPs are roughly in linkage equilibrium) before applying our method of heritability estimation. One approach is to employ a series of $r^2$ thresholds (e.g., decreasing $r^2$ from 0.1 to 0.001) and consider the point at which heritability became stable. Our empirical applications showed that an $r^2$ threshold of ~0.01 may be sufficient. The resulting SNP-based heritability may be considered a conservative estimate (due to possibility of removing some causal variants during LD-pruning). While not directly modeling LD is a limitation of this approach, the lower reliance on accurate LD information may be advantageous in some cases (e.g. when in-sample LD information is not available and only limited external reference data is present).

We have not investigated methods for SE estimation when raw genotype data are available. When raw data are available, one potential approach is to simply resample the individuals with replacement (i.e., standard non-parametric bootstrap). However, such an approach is computationally intensive and the performance over methods based on summary statistics requires further research.

The above resampling methods can potentially be sped up by splitting the job into multiple processes to be run in parallel, although this approach has not been implemented in our software yet. We have not yet fully evaluated the building of confidence interval (CI) in our study, but a natural approach is to assume normality and calculate CI in the form of $\hat{\theta} \pm 1.96 SE$. Assuming a polygenic model, the total heritability is the sum of Vg contributed by many variants of small to modest effect sizes. It is hence reasonable to assume normality by the central limit theorem (as is assumed by other heritability estimation tools). Further research may focus on developing other methods of building CIs and their comparisons.

Here we have also applied our approach to estimate the heritability of different cytokines, which play important roles in immune response and the pathogenesis of autoimmune, inflammatory and infectious diseases. Our analyses suggest that the studied cytokines are moderately heritable in general.

To summarize, SumVg is useful for triangulating evidence from different approaches to support conclusions regarding SNP-based heritability. We present novel methods of computing SE and an easy-to-use software here, which we believe will be helpful to other researchers. Our application to cytokine levels also shed light on the genetic architecture of these clinically important immune traits.

**Acknowledgements**

This study was partially supported by an NSFC grant (81971706), a Collaborative Research Fund (CRF) (C4054-17W) and a Theme-based Research Grant (T44-410/21-N) from the Research Grants Council. HCS was also supported by the Lo Kwee Seong Biomedical Research Fund. We would also like to thank Ms Jinghong QIU for the help in formatting the manuscript.

**Competing interests**

The authors declared no competing interests.

Table 1. Standard error (SE) of the sum of variance explained estimated by different resampling approaches

| Sum_of_Vg | Sample_size | Mean_Est | TRUE_SE | SE | | | | |
|---|---|---|---|---|---|---|---|---|
| | | | | jack_del_1 | jack_del_d | paraboot | fdrboot1 | fdrboot2 |
| 0.295 | 5000 | 0.232 | 0.0482 | 0.0672 | 0.0524 | 0.0488 | 0.0519 | 0.0489 |
| | 10000 | 0.210 | 0.0265 | 0.0353 | 0.0295 | 0.0285 | 0.0312 | 0.0287 |
| | 20000 | 0.244 | 0.0158 | 0.0208 | 0.0185 | 0.0165 | 0.0156 | 0.0168 |
| | 50000 | 0.283 | 0.0076 | 0.0149 | 0.0167 | 0.0063 | 0.0081 | 0.0063 |
| | 1.00E+05 | 0.312 | 0.0063 | 0.0143 | 0.0172 | 0.0051 | 0.0055 | 0.0054 |
| | 2.00E+05 | 0.321 | 0.0045 | 0.0134 | 0.0161 | 0.0036 | 0.0038 | 0.0041 |
| 0.191 | 5000 | 0.207 | 0.0491 | 0.0706 | 0.0523 | 0.0486 | 0.0500 | 0.0485 |
| | 10000 | 0.147 | 0.0242 | 0.0357 | 0.0274 | 0.0263 | 0.0285 | 0.0265 |
| | 20000 | 0.158 | 0.0159 | 0.0208 | 0.0166 | 0.0156 | 0.0162 | 0.0160 |
| | 50000 | 0.174 | 0.0064 | 0.0113 | 0.0113 | 0.0061 | 0.0070 | 0.0061 |
| | 1.00E+05 | 0.195 | 0.0045 | 0.0110 | 0.0131 | 0.0040 | 0.0047 | 0.0041 |
| | 2.00E+05 | 0.207 | 0.0035 | 0.0103 | 0.0128 | 0.0031 | 0.0034 | 0.0035 |
| 0.101 | 5000 | 0.197 | 0.0521 | 0.0692 | 0.0524 | 0.0484 | 0.0496 | 0.0483 |
| | 10000 | 0.116 | 0.0260 | 0.0345 | 0.0265 | 0.0251 | 0.0257 | 0.0251 |
| | 20000 | 0.098 | 0.0143 | 0.0202 | 0.0159 | 0.0150 | 0.0153 | 0.0154 |
| | 50000 | 0.091 | 0.0058 | 0.0098 | 0.0078 | 0.0063 | 0.0057 | 0.0063 |
| | 1.00E+05 | 0.094 | 0.0032 | 0.0069 | 0.0076 | 0.0027 | 0.0036 | 0.0027 |
| | 2.00E+05 | 0.107 | 0.0028 | 0.0072 | 0.0083 | 0.0023 | 0.0027 | 0.0025 |

Sum_of_Vg, true total heritability explained;

Mean_Est, mean estimated variance explained based on corrected z-statistics;

True_SE, SE by repeating experiments 100 times;

jack_del_1, delete-1-jackknife;

jack_del_d, delete-d-jackknife with d equal to 20000;

paraboot, parametric bootstrap approach as described in the text;

fdrboot1, a "weighted" bootstrap approach with consideration of the local fdr, using the observed z-statistic as the mean in each simulation;

fdrboot2, a "weighted" bootstrap approach with consideration of the local fdr, using the corrected z-statistic as the mean in each simulation;

Table 2. Bias, Variance and Root mean squared error (RMSE) of SE for the sum of variance explained estimated by different resampling approaches

| Sum_Vg | N | Bias of the estimator | | | | | Variance of the estimator | | | | | RMSE of the estimator | | | | |
|---|---|---|---|---|---|---|---|---|---|---|---|---|---|---|---|---|
| | | jack_del_1 | jack_del_d | paraboot | fdrboot1 | fdrboot2 | jack_del_1 | jack_del_d | paraboot | fdrboot1 | fdrboot2 | jack_del_1 | jack_del_d | paraboot | fdrboot1 | fdrboot2 |
| 0.295 | 5000 | 1.91E-02 | 4.26E-03 | **5.92E-04** | 3.73E-03 | 7.16E-04 | 1.77E-04 | 5.14E-05 | 1.26E-05 | 1.15E-05 | **9.30E-06** | 2.32E-02 | 8.34E-03 | 3.59E-03 | 5.04E-03 | **3.13E-03** |
| | 10000 | 8.81E-03 | 2.98E-03 | **2.00E-03** | 4.66E-03 | 2.25E-03 | 7.34E-05 | 1.21E-05 | 3.87E-06 | 4.99E-06 | **3.62E-06** | 1.23E-02 | 4.58E-03 | **2.80E-03** | 5.17E-03 | 2.94E-03 |
| | 20000 | 5.04E-03 | 2.78E-03 | 7.37E-04 | -1.46E-04 | 1.07E-03 | 9.06E-05 | 2.27E-06 | **8.33E-07** | 1.12E-06 | 1.00E-06 | 1.08E-02 | 3.16E-03 | 1.17E-03 | **1.07E-03** | 1.47E-03 |
| | 50000 | 7.25E-03 | 9.03E-03 | -1.32E-03 | **4.68E-04** | -1.30E-03 | 1.29E-04 | 1.45E-06 | 1.36E-07 | 1.93E-07 | **1.30E-07** | 1.35E-02 | 9.11E-03 | 1.37E-03 | **6.42E-04** | 1.35E-03 |
| | 1.00E+05 | 7.97E-03 | 1.09E-02 | -1.20E-03 | -8.78E-04 | -9.34E-04 | 1.41E-04 | 1.52E-06 | **8.11E-08** | 3.48E-07 | 1.00E-07 | 1.43E-02 | 1.10E-02 | 1.23E-03 | 1.06E-03 | **9.86E-04** |
| | 2.00E+05 | 8.92E-03 | 1.16E-02 | -8.57E-04 | -6.32E-04 | -3.72E-04 | 1.37E-04 | 8.70E-07 | **3.49E-08** | 1.30E-07 | 4.06E-08 | 1.47E-02 | 1.16E-02 | 8.77E-04 | 7.27E-04 | **4.23E-04** |
| 0.191 | 5000 | 2.16E-02 | 3.21E-03 | -5.02E-04 | 9.69E-04 | -5.23E-04 | 5.53E-04 | 5.41E-05 | 1.07E-05 | 1.02E-05 | **8.58E-06** | 3.19E-02 | 8.03E-03 | 3.31E-03 | 3.34E-03 | **2.98E-03** |
| | 10000 | 1.15E-02 | 3.16E-03 | **2.04E-03** | 4.22E-03 | 2.29E-03 | 2.80E-04 | 1.43E-05 | 4.88E-06 | **2.98E-06** | 5.03E-06 | 2.03E-02 | 4.93E-03 | **3.01E-03** | 4.56E-03 | 3.20E-03 |
| | 20000 | 4.96E-03 | 7.22E-04 | -2.41E-04 | 3.15E-04 | **9.85E-05** | 1.19E-04 | 2.70E-06 | 1.09E-06 | 1.31E-06 | **7.64E-07** | 1.20E-02 | 1.79E-03 | 1.07E-03 | 1.19E-03 | **8.80E-04** |
| | 50000 | 4.90E-03 | 4.83E-03 | -2.92E-04 | 5.76E-04 | -2.97E-04 | 1.10E-04 | 8.17E-07 | 1.66E-07 | 1.50E-07 | **1.24E-07** | 1.16E-02 | 4.91E-03 | 5.01E-04 | 6.94E-04 | **4.60E-04** |
| | 1.00E+05 | 6.45E-03 | 8.56E-03 | -5.40E-04 | **1.30E-04** | -4.33E-04 | 1.32E-04 | 1.68E-06 | **6.30E-08** | 1.32E-07 | 8.37E-08 | 1.32E-02 | 8.65E-03 | 5.96E-04 | **3.85E-04** | 5.21E-04 |
| | 2.00E+05 | 6.85E-03 | 9.28E-03 | -3.57E-04 | -1.41E-04 | -3.31E-05 | 1.28E-04 | 1.06E-06 | **3.04E-08** | 1.54E-07 | 3.78E-08 | 1.32E-02 | 9.34E-03 | 3.97E-04 | 4.17E-04 | **1.97E-04** |
| 0.101 | 5000 | 1.71E-02 | **3.38E-04** | -3.72E-03 | -2.54E-03 | -3.81E-03 | 1.81E-04 | 6.26E-05 | 1.01E-05 | 1.07E-05 | **7.70E-06** | 2.17E-02 | 7.92E-03 | 4.90E-03 | **4.14E-03** | 4.71E-03 |
| | 10000 | 8.45E-03 | 4.62E-04 | -8.81E-04 | -2.84E-04 | -9.03E-04 | 1.66E-04 | 1.45E-05 | 3.49E-06 | **1.66E-06** | 2.46E-06 | 1.54E-02 | 3.83E-03 | 2.07E-03 | **1.32E-03** | 1.81E-03 |
| | 20000 | 5.92E-03 | 1.56E-03 | 7.21E-04 | 1.04E-03 | 1.08E-03 | 1.44E-04 | 4.20E-06 | 8.80E-07 | 1.22E-06 | **8.17E-07** | 1.34E-02 | 2.57E-03 | **1.18E-03** | 1.52E-03 | 1.41E-03 |
| | 50000 | 4.03E-03 | 2.04E-03 | 4.91E-04 | -1.02E-04 | 5.85E-04 | 6.00E-05 | 4.31E-07 | 1.61E-07 | **1.08E-07** | 1.26E-07 | 8.73E-03 | 2.15E-03 | 6.34E-04 | **3.44E-04** | 6.85E-04 |
| | 1.00E+05 | 3.73E-03 | 4.34E-03 | -5.10E-04 | **4.31E-04** | -5.17E-04 | 9.13E-05 | 3.03E-07 | 2.83E-08 | 4.34E-08 | **2.05E-08** | 1.03E-02 | 4.38E-03 | 5.37E-04 | **4.79E-04** | 5.36E-04 |
| | 2.00E+05 | 4.38E-03 | 5.48E-03 | -5.31E-04 | -1.61E-04 | -3.81E-04 | 1.00E-04 | 3.41E-07 | **2.22E-08** | 7.06E-08 | 2.88E-08 | 1.09E-02 | 5.51E-03 | 5.52E-04 | **3.11E-04** | 4.17E-04 |

Legend for table 2:

Sum_Vg, true total heritability explained; N, sample size. The best performing method (for estimation of SE) in each scenario is in bold.

For other abbreviations, please refer to table 1.

Table 3. Summary of the immune traits being studied

| Trait | Abbreviation | GWAS ID | N | SNP_h2 (LDSC) | SNP_h2_se (LDSC) |
|---|---|---|---|---|---|
| Stem cell factor | SCF | ebi-a-GCST004429 | 8290 | -0.06 | 0.055 |
| Interleukin-4 | IL4 | ebi-a-GCST004453 | 8124 | -0.0446 | 0.0595 |
| Interleukin-17 | IL17 | ebi-a-GCST004442 | 7760 | -0.0407 | 0.0623 |
| Hepatocyte growth factor | HGF | ebi-a-GCST004449 | 8292 | -0.0311 | 0.0579 |
| Basic fibroblast growth factor | FGFBasic | ebi-a-GCST004459 | 7565 | -0.0159 | 0.0597 |
| Stromal cell-derived factor-1 alpha (CXCL12) | SDF1a | ebi-a-GCST004427 | 5998 | -0.0116 | 0.0713 |
| Interleukin-6 | IL6 | ebi-a-GCST004446 | 8189 | -0.0071 | 0.0568 |
| Platelet derived growth factor BB | PDGFbb | ebi-a-GCST004432 | 8293 | -0.0043 | 0.0624 |
| TNF-related apoptosis inducing ligand | TRAIL | ebi-a-GCST004424 | 8186 | 0.0125 | 0.0613 |
| Interferon-gamma | IFNg | ebi-a-GCST004456 | 7701 | 0.0134 | 0.0624 |
| Granulocyte colony-stimulating factor | GCSF | ebi-a-GCST004458 | 7904 | 0.0173 | 0.0601 |
| Interleukin-10 | IL10 | ebi-a-GCST004444 | 7681 | 0.0186 | 0.0691 |

Trait, trait name of analyzed GWAS dataset; Abbreviation, abbreviation of the trait name;

GWAS ID, ID of GWAS dataset for downloading from the IEU OpenGWAS Project;

N, sample size;

SNP_h2 (LDSC), SNP heritability estimated by LDSC as reported in GWASAtlas;

SNP_h2_se (LDSC), standard error of SNP heritability estimated by LDSC as reported in GWASAtlas;

Table 4. SE of the sum of variance explained estimated by different resampling approaches for 12 immune traits

| Trait | N | LDSC | | SumVg | | | | | | | |
|---|---|---|---|---|---|---|---|---|---|---|---|
| | | h2 | se | h2 | r2 | n_pruned_snp | se_jack1 | se_jack_del_d | se_paraboot | se_fdrboot1 | se_fdrboot2 |
| SCF | 8290 | -0.06 | 0.055 | 0.333 | 0.1 | 428593 | 0.0926 | 0.0822 | 0.0679 | 0.0443 | 0.0514 |
| | | | | 0.185 | 0.05 | 251008 | 0.0526 | 0.0456 | 0.0467 | 0.0502 | 0.0517 |
| | | | | 0.105 | 0.025 | 127908 | 0.0307 | 0.0313 | 0.0272 | 0.0397 | 0.0335 |
| | | | | **0.100** | **0.01** | **61938** | **0.0310** | **0.0200** | **0.0220** | **0.0252** | **0.0265** |
| | | | | 0.092 | 0.005 | 51370 | 0.0229 | 0.0169 | 0.0201 | 0.0235 | 0.0230 |
| | | | | 0.101 | 0.002 | 48088 | 0.0319 | 0.0153 | 0.0220 | 0.0226 | 0.0198 |
| | | | | 0.102 | 0.001 | 47108 | 0.0316 | 0.0155 | 0.0223 | 0.0216 | 0.0188 |
| IL4 | 8124 | -0.0446 | 0.0595 | 0.503 | 0.1 | 427005 | 0.1218 | 0.1133 | 0.0616 | 0.0563 | 0.0569 |
| | | | | 0.377 | 0.05 | 249710 | 0.1000 | 0.0823 | 0.0484 | 0.0445 | 0.0453 |
| | | | | 0.302 | 0.025 | 127248 | 0.0650 | 0.0594 | 0.0318 | 0.0365 | 0.0336 |
| | | | | **0.235** | **0.01** | **61685** | **0.0529** | **0.0313** | **0.0247** | **0.0240** | **0.0236** |
| | | | | 0.215 | 0.005 | 51196 | 0.0472 | 0.0278 | 0.0227 | 0.0217 | 0.0221 |
| | | | | 0.197 | 0.002 | 47878 | 0.0571 | 0.0273 | 0.0228 | 0.0225 | 0.0253 |
| | | | | 0.187 | 0.001 | 46911 | 0.0482 | 0.0244 | 0.0198 | 0.0226 | 0.0242 |
| IL17 | 7760 | -0.0407 | 0.0623 | 0.352 | 0.1 | 427226 | 0.1240 | 0.0946 | 0.0692 | 0.0625 | 0.0609 |
| | | | | 0.228 | 0.05 | 250259 | 0.0683 | 0.0668 | 0.0499 | 0.0495 | 0.0495 |
| | | | | 0.299 | 0.025 | 127479 | 0.0877 | 0.0568 | 0.0360 | 0.0380 | 0.0323 |
| | | | | **0.234** | **0.01** | **61756** | **0.0485** | **0.0340** | **0.0239** | **0.0267** | **0.0256** |
| | | | | 0.196 | 0.005 | 51215 | 0.0475 | 0.0295 | 0.0237 | 0.0190 | 0.0249 |
| | | | | 0.195 | 0.002 | 47887 | 0.0634 | 0.0231 | 0.0231 | 0.0226 | 0.0210 |
| | | | | 0.188 | 0.001 | 46931 | 0.0568 | 0.0242 | 0.0183 | 0.0211 | 0.0215 |
| HGF | 8292 | -0.0311 | 0.0579 | 0.366 | 0.1 | 428318 | 0.0917 | 0.0864 | 0.0569 | 0.0642 | 0.0593 |
| | | | | 0.242 | 0.05 | 250843 | 0.0812 | 0.0722 | 0.0483 | 0.0492 | 0.0491 |
| | | | | 0.205 | 0.025 | 127850 | 0.0657 | 0.0488 | 0.0327 | 0.0326 | 0.0357 |

| | | | | 0.098 | 0.01 | 61906 | 0.0379 | 0.0224 | 0.0225 | 0.0260 | 0.0242 |
| --- | --- | --- | --- | --- | --- | --- | --- | --- | --- | --- | --- |
| | | | | 0.115 | 0.005 | 51301 | 0.0347 | 0.0199 | 0.0224 | 0.0230 | 0.0203 |
| | | | | 0.111 | 0.002 | 47878 | 0.0414 | 0.0162 | 0.0189 | 0.0211 | 0.0215 |
| | | | | 0.108 | 0.001 | 46934 | 0.0312 | 0.0171 | 0.0221 | 0.0208 | 0.0211 |
| FGFBasic | 7565 | -0.0159 | 0.0597 | 0.269 | 0.1 | 427284 | 0.0835 | 0.0902 | 0.0656 | 0.0530 | 0.0577 |
| | | | | 0.217 | 0.05 | 249930 | 0.0891 | 0.0604 | 0.0473 | 0.0504 | 0.0468 |
| | | | | 0.117 | 0.025 | 127587 | 0.0452 | 0.0431 | 0.0340 | 0.0363 | 0.0358 |
| | | | | 0.133 | 0.01 | 61911 | 0.0408 | 0.0301 | 0.0232 | 0.0239 | 0.0275 |
| | | | | 0.135 | 0.005 | 51259 | 0.0376 | 0.0243 | 0.0242 | 0.0267 | 0.0219 |
| | | | | 0.143 | 0.002 | 47874 | 0.0362 | 0.0218 | 0.0185 | 0.0233 | 0.0245 |
| | | | | 0.126 | 0.001 | 46914 | 0.0392 | 0.0206 | 0.0227 | 0.0214 | 0.0208 |
| SDF1a | 5998 | -0.0116 | 0.0713 | 0.395 | 0.1 | 425165 | 0.1120 | 0.1068 | 0.0731 | 0.0757 | 0.0870 |
| | | | | 0.256 | 0.05 | 248727 | 0.0872 | 0.0750 | 0.0580 | 0.0565 | 0.0631 |
| | | | | 0.213 | 0.025 | 126986 | 0.0707 | 0.0462 | 0.0431 | 0.0472 | 0.0468 |
| | | | | 0.163 | 0.01 | 61680 | 0.0497 | 0.0380 | 0.0359 | 0.0349 | 0.0324 |
| | | | | 0.190 | 0.005 | 51092 | 0.0708 | 0.0318 | 0.0250 | 0.0297 | 0.0270 |
| | | | | 0.165 | 0.002 | 47702 | 0.0447 | 0.0270 | 0.0294 | 0.0304 | 0.0301 |
| | | | | 0.159 | 0.001 | 46789 | 0.0512 | 0.0232 | 0.0279 | 0.0258 | 0.0308 |
| IL6 | 8189 | -0.0071 | 0.0568 | 0.422 | 0.1 | 427566 | 0.0878 | 0.0896 | 0.0510 | 0.0575 | 0.0594 |
| | | | | 0.227 | 0.05 | 250247 | 0.0620 | 0.0713 | 0.0402 | 0.0463 | 0.0468 |
| | | | | 0.158 | 0.025 | 127503 | 0.0672 | 0.0457 | 0.0372 | 0.0300 | 0.0360 |
| | | | | 0.139 | 0.01 | 61931 | 0.0606 | 0.0258 | 0.0220 | 0.0247 | 0.0220 |
| | | | | 0.114 | 0.005 | 51332 | 0.0288 | 0.0176 | 0.0196 | 0.0227 | 0.0236 |
| | | | | 0.115 | 0.002 | 47930 | 0.0302 | 0.0164 | 0.0191 | 0.0226 | 0.0202 |
| | | | | 0.117 | 0.001 | 46944 | 0.0319 | 0.0175 | 0.0227 | 0.0209 | 0.0211 |
| PDGFbb | 8293 | -0.0043 | 0.0624 | 0.432 | 0.1 | 427743 | 0.0907 | 0.0993 | 0.0726 | 0.0653 | 0.0676 |
| | | | | 0.341 | 0.05 | 250325 | 0.0670 | 0.0808 | 0.0600 | 0.0496 | 0.0576 |
| | | | | 0.307 | 0.025 | 127567 | 0.0735 | 0.0554 | 0.0370 | 0.0334 | 0.0326 |

| | | | | | | | | | | |
|---|---|---|---|---|---|---|---|---|---|---|
| | | | | **0.154** | **0.01** | **61789** | **0.0372** | **0.0250** | **0.0213** | **0.0245** | **0.0243** |
| | | | | 0.125 | 0.005 | 51140 | 0.0310 | 0.0226 | 0.0234 | 0.0230 | 0.0221 |
| | | | | 0.120 | 0.002 | 47822 | 0.0258 | 0.0205 | 0.0214 | 0.0208 | 0.0233 |
| | | | | 0.117 | 0.001 | 46853 | 0.0392 | 0.0192 | 0.0201 | 0.0209 | 0.0226 |
| TRAIL | 8186 | 0.0125 | 0.0613 | 0.559 | 0.1 | 423391 | 0.0613 | 0.1018 | 0.0785 | 0.0790 | 0.0750 |
| | | | | 0.304 | 0.05 | 247717 | 0.0543 | 0.1190 | 0.0526 | 0.0503 | 0.0439 |
| | | | | 0.242 | 0.025 | 126350 | 0.0607 | 0.0647 | 0.0321 | 0.0362 | 0.0370 |
| | | | | **0.128** | **0.01** | **61114** | **0.0316** | **0.0251** | **0.0242** | **0.0229** | **0.0277** |
| | | | | 0.127 | 0.005 | 50633 | 0.0298 | 0.0231 | 0.0255 | 0.0239 | 0.0268 |
| | | | | 0.128 | 0.002 | 47359 | 0.0332 | 0.0216 | 0.0233 | 0.0215 | 0.0266 |
| | | | | 0.121 | 0.001 | 46415 | 0.0358 | 0.0195 | 0.0222 | 0.0229 | 0.0256 |
| IFNg | 7701 | 0.0134 | 0.0624 | 0.393 | 0.1 | 426740 | 0.0946 | 0.0811 | 0.0528 | 0.0590 | 0.0594 |
| | | | | 0.241 | 0.05 | 249818 | 0.0655 | 0.0628 | 0.0553 | 0.0520 | 0.0509 |
| | | | | 0.244 | 0.025 | 127514 | 0.0734 | 0.0582 | 0.0330 | 0.0406 | 0.0320 |
| | | | | **0.138** | **0.01** | **61890** | **0.0289** | **0.0303** | **0.0267** | **0.0239** | **0.0257** |
| | | | | 0.138 | 0.005 | 51314 | 0.0424 | 0.0201 | 0.0222 | 0.0248 | 0.0293 |
| | | | | 0.141 | 0.002 | 47918 | 0.0321 | 0.0204 | 0.0251 | 0.0248 | 0.0286 |
| | | | | 0.137 | 0.001 | 46934 | 0.0253 | 0.0183 | 0.0223 | 0.0246 | 0.0233 |
| GCSF | 7904 | 0.0173 | 0.0601 | 0.246 | 0.1 | 427393 | 0.0707 | 0.0820 | 0.0620 | 0.0604 | 0.0580 |
| | | | | 0.198 | 0.05 | 250222 | 0.0636 | 0.0607 | 0.0402 | 0.0436 | 0.0486 |
| | | | | 0.164 | 0.025 | 127583 | 0.0501 | 0.0415 | 0.0302 | 0.0360 | 0.0327 |
| | | | | **0.142** | **0.01** | **61846** | **0.0434** | **0.0257** | **0.0280** | **0.0239** | **0.0257** |
| | | | | 0.122 | 0.005 | 51266 | 0.0379 | 0.0196 | 0.0205 | 0.0247 | 0.0238 |
| | | | | 0.120 | 0.002 | 47919 | 0.0413 | 0.0183 | 0.0219 | 0.0236 | 0.0201 |
| | | | | 0.112 | 0.001 | 46939 | 0.0312 | 0.0159 | 0.0234 | 0.0207 | 0.0202 |
| IL10 | 7681 | 0.0186 | 0.0691 | 0.331 | 0.1 | 427218 | 0.0621 | 0.1019 | 0.0584 | NA | NA |
| | | | | 0.310 | 0.05 | 250109 | 0.0670 | 0.0858 | 0.0448 | NA | NA |
| | | | | 0.198 | 0.025 | 127543 | 0.0566 | 0.0463 | 0.0356 | 0.0382 | 0.0406 |

| | | | | | | | |
|---|---|---|---|---|---|---|---|
| **0.130** | **0.01** | **61944** | **0.0328** | **0.0225** | **0.0251** | **0.0268** | **0.0258** |
| 0.141 | 0.005 | 51257 | 0.0400 | 0.0220 | 0.0237 | 0.0282 | 0.0238 |
| 0.148 | 0.002 | 47880 | 0.0433 | 0.0183 | 0.0204 | 0.0271 | 0.0231 |
| 0.142 | 0.001 | 46898 | 0.0317 | 0.0194 | 0.0219 | 0.0261 | 0.0228 |

Trait, N, LDSC(h2, se) are the same meaning as in Table 3;

h2, heritability estimated by SumVg across a set of $r^2$ thresholds;

r2, the r2 pruning threshold;

n_pruned_snp, number of SNPs after LD pruning at the corresponding $r^2$ threshold;

se_jack_1, se_jack_del_d, se_paraboot, se_fdrboot1 and se_fdrboot2 are SE estimated by different approaches as we described above;

"NA" was denoted when "locfdr" failed to estimate local false discovery rate.

The estimates with r2=0.01 were highlighted, as we observed that in general the heritability estimates stabilize at r2 ~ 0.01.

Figure 1 Boxplot of SE estimated by different approaches

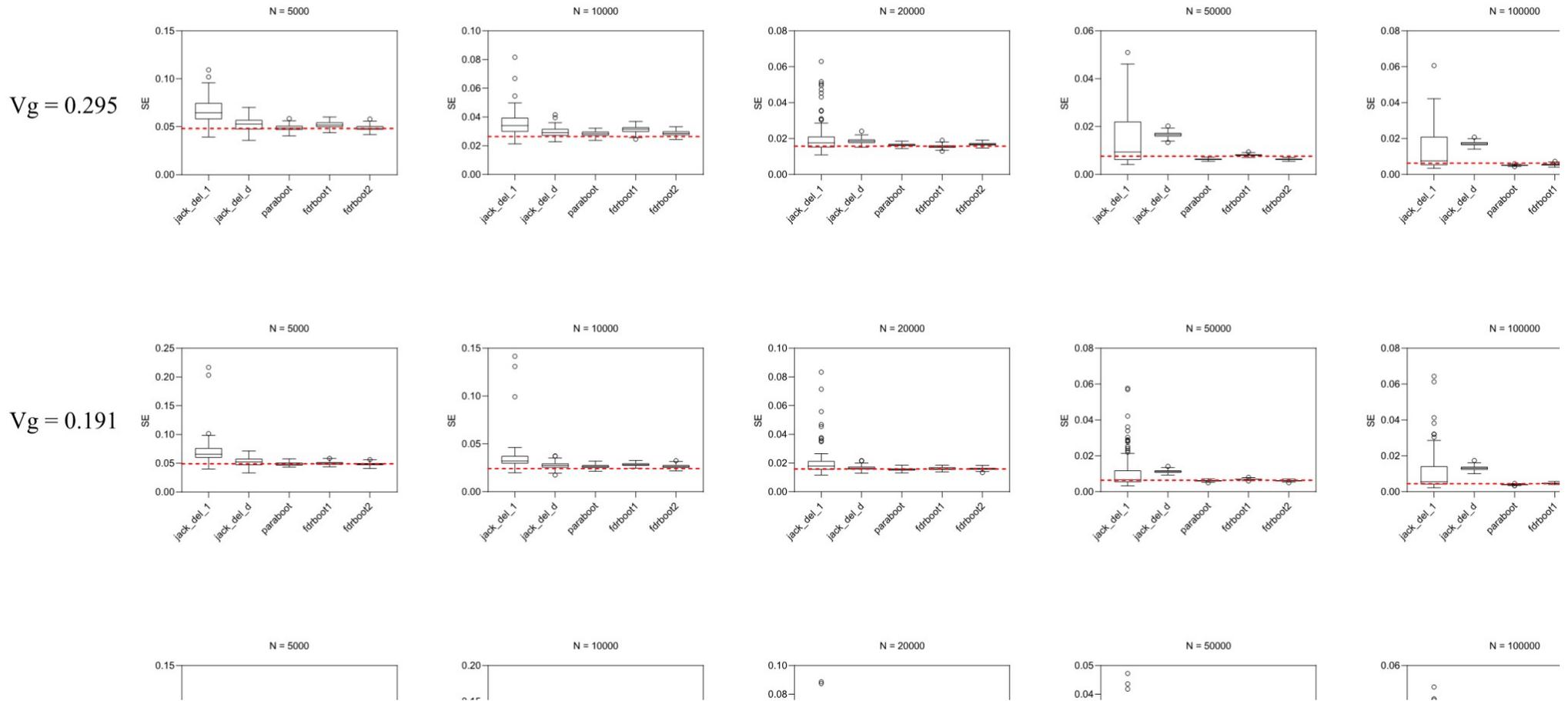

Vg is the sum of variance explained, N is the sample size, and the horizontal line refers to true SE calculated by repeating experiments 100 times. jack_del_1, jack_del_d, paraboot, fdrboot1 and fdrboot2 are different SE estimation approaches as described above.

Figure 2 Bar plot of Bias, Variance and RMSE of SE estimated by different approaches

$Vg = 0.295$ 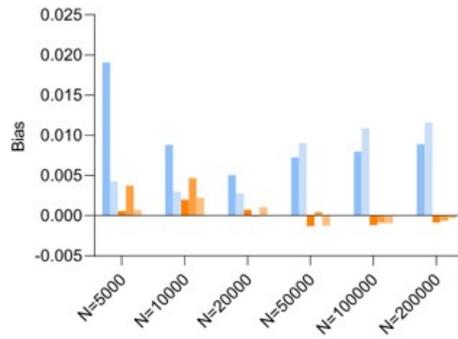 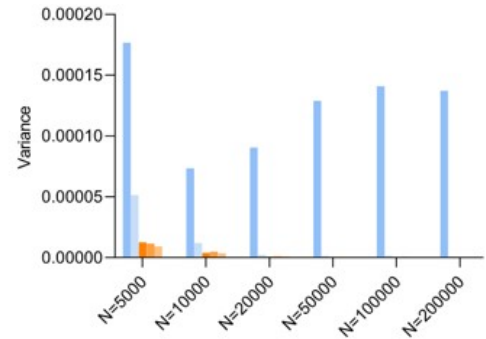 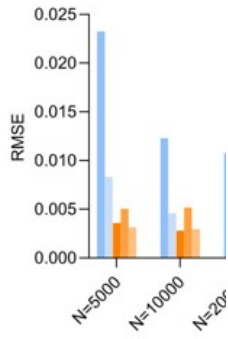

$Vg = 0.191$ 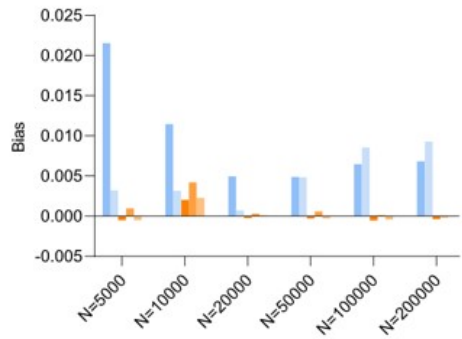 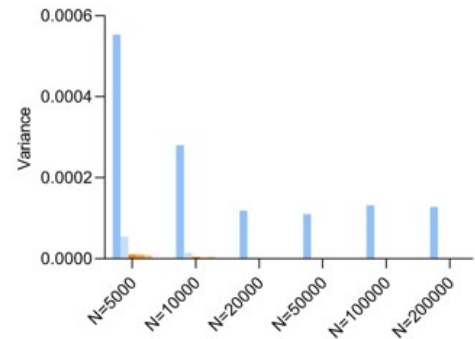 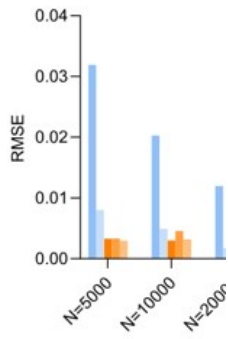

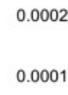

Vg is the sum of variance explained, and N is the sample size. jack_del_1, jack_del_d, paraboot, fdrboot1 and fdrboot2 are different SE estimation approaches as described above.